\documentclass[letterpaper]{article} 
\usepackage{aaai25}  
\usepackage{times}  
\usepackage{helvet}  
\usepackage{courier}  
\usepackage[hyphens]{url}  
\usepackage{graphicx} 
\usepackage{natbib}  
\usepackage{caption} 
\usepackage{algorithm}
\usepackage{algorithmic}

\usepackage{url}            
\usepackage{booktabs}       
\usepackage{amsfonts}       

\usepackage{microtype}      
\usepackage{xcolor}         
\usepackage{xspace}
\usepackage{amsmath}
\usepackage{multirow}
\usepackage{arydshln} 
\usepackage{pgfplots}
\usepackage{enumitem}
\usepackage{subcaption}
\newcommand{\tool}{EvalSVA\xspace}
\newcommand{\baseline}{Single\xspace}
\newcommand{\wxc}[1]{\textcolor{brown}{{#1}}}

\definecolor{darkgreen}{RGB}{0, 200, 0}

\usepackage{newfloat}
\usepackage{listings}

\DeclareCaptionStyle{ruled}{labelfont=normalfont,labelsep=colon,strut=off} 
\floatstyle{ruled}
\newfloat{listing}{tb}{lst}{}
\floatname{listing}{Listing}
\title{\tool: Multi-Agent Evaluators for Next-Gen Software Vulnerability Assessment}
\author{
    Xin-Cheng Wen\textsuperscript{\rm 1}, Jiaxin Ye\textsuperscript{\rm 2}, Cuiyun Gao\textsuperscript{\rm 1,\rm 3}\thanks{The corresponding author.}, Lianwei Wu\textsuperscript{\rm 4}, Qing Liao\textsuperscript{\rm 1,\rm 3},\\
}
\affiliations{
    \textsuperscript{\rm 1}Harbin Institute of Technology, Shenzhen, China\\
    \textsuperscript{\rm 2}Fudan University, Shanghai, China\\
    \textsuperscript{\rm 3}Peng Cheng Laboratory, Shenzhen, China\\
    \textsuperscript{\rm 4}Northwestern Polytechnical University, Xi'an, China\\

    xiamenwxc@foxmail.com, jxye22@m.fudan.edu.cn, gaocuiyun@hit.edu.cn, wlw@nwpu.edu.cn, liaoqing@hit.edu.cn
}

\usepackage{bibentry}

\begin{document}

\maketitle

\begin{abstract}
Software Vulnerability (SV) assessment is a crucial process of determining different aspects of SVs (e.g., attack vectors and scope) for developers to effectively prioritize efforts in vulnerability mitigation. 
It presents a challenging and laborious process due to the complexity of SVs and the scarcity of labeled data.
To mitigate the above challenges, we introduce \tool, a multi-agent evaluators team to autonomously deliberate and evaluate various aspects of SV assessment. 
Specifically, we propose a multi-agent-based framework to simulate vulnerability assessment strategies in real-world scenarios, which employs multiple Large Language Models (LLMs) into an integrated group to enhance the effectiveness of SV assessment in the limited data. 
We also design diverse communication strategies to autonomously discuss and assess different aspects of SV. 
Furthermore, we construct a multi-lingual SV assessment dataset based on the new standard of CVSS, comprising 699, 888, and 1,310 vulnerability-related commits in C++, Python, and Java, respectively. 
Our experimental results demonstrate that \tool averagely outperforms the 44.12\% accuracy and 43.29\% F1 for SV assessment compared with the previous methods. It shows that \tool offers a human-like process and generates both reason and answer for SV assessment. 
\tool can also aid human experts in SV assessment, which provides more explanation and details for SV assessment.
\label{abstract}
\end{abstract}

%

\section{Introduction}
\label{introduction}
Software Vulnerabilities (SVs) are mostly caused by insecure code that can be exploited to attack software systems~\cite{DBLP:conf/kbse/DissanayakeJZB22, DBLP:series/ccn/KhanP18}, and further cause security issues such as systems susceptible to cyber-attacks, and data leakage problems~\cite{DBLP:journals/csur/LeCB23}.  
In the past ten years, the number of SVs has been increasing rapidly~\cite{DBLP:journals/ns/Smyth17}, rising from 5,697 in 2013 to 29,065 in 2023~\cite{Statista1}.  Therefore, SV assessment is a crucial yet challenging problem in security.

The expert-based Common Vulnerability Scoring System (CVSS)~\cite{CVSS} is a widely adopted framework for assessing SVs. CVSS provides metrics to quantify the exploitability, impact, and severity metrics of SVs~\cite{CVSSmetrics, Park}. 
Such procedures are labor-intensive and suffer from inefficiencies due to 
the complexity of vulnerabilities~\cite{DBLP:conf/ccs/BilgeD12, DBLP:conf/ic-nc/FeutrillRYR18}.
Traditional automated approaches for SV assessment, primarily reliant on user-submitted SV reports, are hampered by substantial delays—over 82\% of reports are filed more than 30 days post initial detection~\cite{DBLP:conf/icsm/ThungLJLRD12}. 
Recent studies aim to automate assess SV via commits~\cite{DBLP:conf/kbse/LeHCB21/DeepCVA, DBLP:conf/kbse/ZhouPW00WH21}, significantly reducing reliance on manual expert evaluations and accelerating the assessment process.

However, the existing methods still pose several major challenges that need to be addressed:
\textit{Firstly}, the existing methods depend on extensive labeled data, which is difficult to evolve in practice. 
Specifically, the CVSS framework updates rapidly, evolving from CVSS v2 to v3, and subsequently to v3.1~\cite{CVSSmetricsV2, CVSSmetricsV3,CVSSmetrics}. It is time-consuming for experts to furnish high-quality assessments in new standards. 
For instance, the National Vulnerability Database (NVD)~\cite{NVD} and the Common Vulnerabilities and Exposures (CVE)~\cite{CVE} lists maintained by Mend~\cite{Mend} only contains 699 complete vulnerability entries for C++ from 2013 to 2023. 
Consequently, the labeled data present difficulties in industry and limit practical value in real-world scenarios, potentially leading to unreliable performance of existing methods.
\textit{Second}, the previous commit-level SV assessment studies have not started to use the new
standards~\cite{CVSSmetrics}, which incorporate additional metrics (e.g., \textit{Scope} and \textit{User Interaction}) to enhance the complexity of vulnerability and become the current standard in
industry. 
\textit{Additionally}, most of the existing techniques 
solely predict SV scores of CVSS. 
They provide no idea about how the vulnerability assessment is derived from the input, making the results difficult to interpret and verify.

\begin{figure*}[t]
	\centering
    \includegraphics[width=0.95\textwidth]{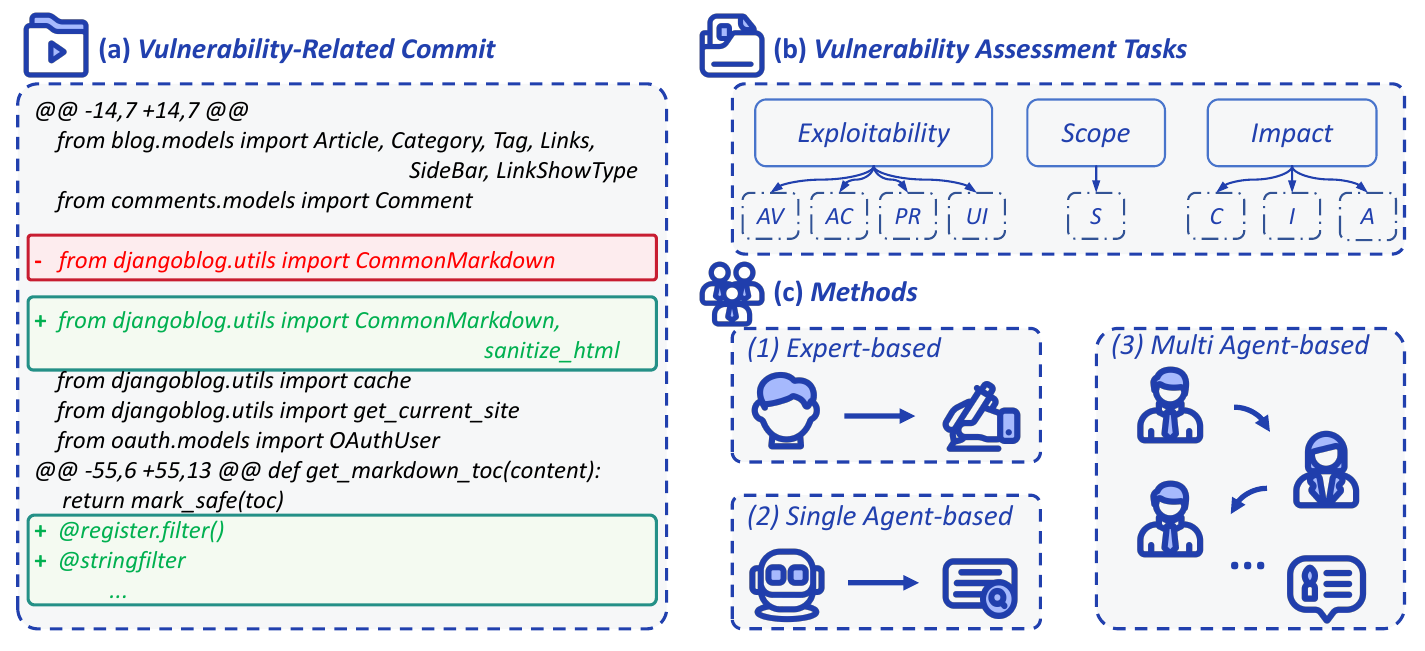}
	\caption{Figure (a) presents the vulnerability-related commit of CVE-2023-2954~\cite{CVE-2023-2954}. The code shaded in \textcolor{red}{red} and \textcolor{darkgreen}{green} denote the \textcolor{red}{vulnerability} code and corresponding \textcolor{darkgreen}{fixed code} from commit, respectively. Figure (b) presents the three aspects and eight tasks of SV assessment. Figure (c) presents the three types of methods for SV assessment.}
	\label{architecture}
\end{figure*}

To mitigate the above challenges, 
we propose a multi-agent \textbf{EVAL}uators team to autonomously deliberate and evaluate various aspects for \textbf{S}oftware \textbf{V}ulnerability \textbf{A}ssessment, called \textbf{\tool}.
Specifically, we propose a multi-agent-based framework to simulate vulnerability assessment strategies in real-world scenarios, which employs multiple Large Language Models (LLMs) into an integrated group to enhance the effectiveness of SV assessment in limited data. 
We also design diverse communication strategies to autonomously discuss, which conduct comprehensive processes and assess different aspects of SV. 
Moreover, to testify the potential of the multi-agent framework in the real-world scenario, we construct the first multi-lingual vulnerability assessment dataset based on the new standard of CVSS, comprising 699, 888, and 1,310 vulnerability-related commits in C++, Python, and Java, respectively. Our case study also shows that \tool offers a human-like process and generates both reason and answer for SV assessment.

In summary, the major contributions of this paper are summarized as follows:
\begin{itemize}[leftmargin=5pt]
\item
We are the first to
propose the multi-agent evaluators with autonomously deliberating for next-gen software vulnerability assessment. Our experimental results demonstrate that \tool averagely outperforms the 44.12\% accuracy and 43.29\% F1 compared with the single agent.

\item We construct the first multi-lingual vulnerability assessment dataset based on the new standard of CVSS, comprising 699, 888, and 1,310 vulnerability-related commits in C++, Python, and Java, respectively.  

\item We explore the performance of different communication strategies. The results show that \tool can aid human experts in many aspects of SV assessment, which provides more explanation for SV assessment and highlights a critical gap in capturing vulnerability patterns.
 

\end{itemize}

\section{Methodology}
In this section, we elaborate on the overview of \tool by first introducing the SV assessment task formulation and then the proposed multi-agent evaluators.

\subsection{Software Vulnerability Assessment Formulation}
\paragraph{Common vulnerability scoring system.}
The CVSS has emerged as the definitive framework for evaluating the severity of SVs. In this paper, we first employ CVSS v3.1 for the commit-level SV assessment.
In this paper, we focus on the prediction of \textit{Base Metrics} due to their broader applicability. These metrics encapsulate the intrinsic attributes of a vulnerability that remain constant over time and across different user environments;

\paragraph{SV assessment task formulation.}
As shown in Figure~\ref{architecture}(a), a vulnerability-related commit can be denoted by the input $\mathcal{X}$ (the template of input $\mathcal{X}$ are shown in Appendix) and SV tasks can be performed in all metrics simultaneously. 
The goal of \tool is to learn a mapping $\mathcal{F}:\mathcal{X}\mapsto\mathcal{Y}$ from input $\mathcal{X}$ to the output signals $\mathcal{Y}$.
Specifically, the output signals for SV assessment tasks can be broadly classified into three aspects: Exploitability, Scope, and Impact.
As shown in Figure~\ref{architecture}(b), the output signals consists of the security of \textit{Attack Vector (AV)} $y_\text{AV}$, \textit{Attack Complexity (AC)} $y_\text{AC}$, \textit{Privileges Required (PR)} $y_\text{PR}$ and \textit{User Interaction (UI)} $y_\text{UI}$ for exploitability aspect, the security of \textit{Scope Change (S)} $y_\text{S}$ for scope aspect, and the security of \textit{Confidentiality (C)} $y_\text{C}$, \textit{Integrity (I)} $y_\text{I}$ and \textit{Availability (A)} $y_\text{A}$ for impact aspect.
We then briefly introduce each task of the SV assessment in the CVSS v3.1 as follows:

\textbf{(1) Exploitability}: 
The exploitability reflects the properties of the vulnerability that lead to a successful attack. In this paper, we use the four metrics to represent the vulnerability exploitability, including \textit{AV}, \textit{AC}, \textit{PR}, and \textit{UI}. 
Specifically, the \textit{AV} metric reflects the attack path by which vulnerability exploitation is possible. 
\textit{AC} metric describes the difficulty of conditions beyond the attacker’s control to exploit the vulnerability.
\textit{PR} metric assesses the level of authority or access rights that an attacker must acquire to successfully exploit the vulnerability.
\textit{UI} metric distinguishes between vulnerabilities that can be exploited solely by attackers and those requiring involvement from a separate user process.
For example, Figure~\ref{architecture}(a)'s original code (shaded in red) contains a Cross-Site Scripting (XSS)~\cite{CWE-79} vulnerability which generally requires ``Low'' privileges, such as a standard user account and ``Required'' user interaction with a potentially victim triggering the malicious script. 
This vulnerability can be exploited through the ``Network'' attack path with ``Low'' complexity via user inputs.

\textbf{(2) Scope}: It indicates whether exploiting a vulnerability impacts resources beyond its security scope (e.g., application and operating system). 
\textit{S} 
determines whether exploiting a vulnerability within a component's scope provides the ability to access or impact the scopes of other components. 

\textbf{(3) Impact}: The impact captures the consequences of a successfully exploited vulnerability, which can cause losses in \textit{Confidentiality}, \textit{Integrity}, and \textit{Availability}. 
\textit{Confidentiality} refers to limiting information access and disclosure while also preventing unauthorized individuals from gaining access. 
\textit{Integrity} refers to the trustworthiness and accuracy of information, ensuring that data remains reliable. 
\textit{Availability} presents the accessibility of information resources, such as processor cycles or disk space.

\begin{figure}[t]
	\centering
    \includegraphics[width=0.5\textwidth]{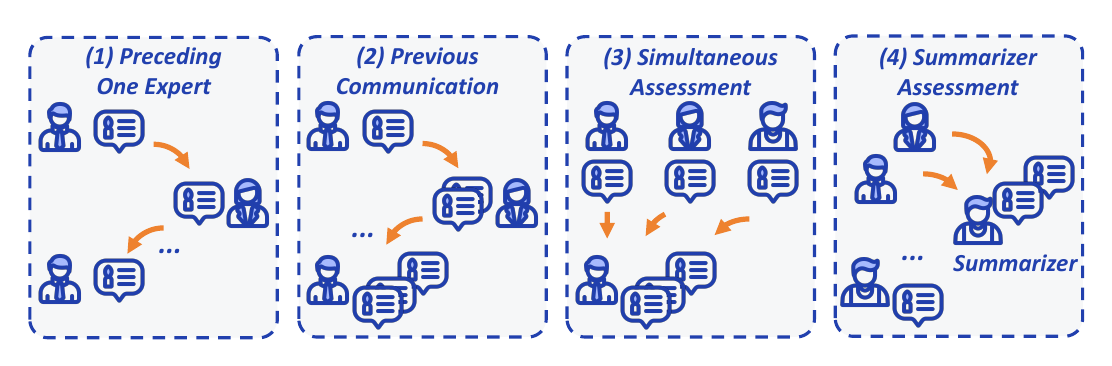}
	\caption{Communication strategy for SV assessment.}
	\label{strategy}
\end{figure}

\subsection{Multi Agent Evaluators}
\subsubsection{Multi Agents and Software Vulnerability Assessment}
Various studies~\cite{DBLP:conf/kbse/GaoWGWZL23, DBLP:conf/kbse/PengWWGL23, DBLP:conf/icse/DengXYZY024} have shown that LLM-based methods are utilized to boost interpretability and practical values behind the classical supervised-based method.
Despite the capability of a single LLM to handle a wide range of tasks across multiple domains~\cite{DBLP:conf/nips/ZhengC00WZL0LXZ23, DBLP:journals/corr/abs-2312-09731}, it continues to encounter significant challenges in SV assessment. 
This is primarily due to assessing the severity of vulnerabilities entails a complex and consequential process~\cite{DBLP:conf/esem/CroftNCB21}, which typically requires collaboration among multiple experts rather than relying solely on individual assessments.
These complex situations make it difficult for an existing single LLM to perform well in SV assessment.
Inspired by the recent advance in multi-agent methods has demonstrated its effectiveness~\cite{DBLP:journals/corr/abs-2303-17760, DBLP:journals/corr/abs-2305-19118, DBLP:journals/corr/abs-2403-11807}, 
we design the first multi-agent-based framework for effectively SV assessment, where the agents interact and communicate within a collaborative environment, aiming to emulate the interaction and collaboration strategies in real-world scenarios~\cite{DBLP:conf/emnlp/KarpinskaAI21}. We elaborate on the two components in \tool including vulnerability expert agents and communication strategy.


\begin{table}[t]
\centering
\setlength{\tabcolsep}{3mm}
\renewcommand{\arraystretch}{1.2}

\caption{Statistics of the dataset in C++, Java and Python.}
\resizebox{0.48\textwidth}{!}{
\begin{tabular}{ccccc}
\toprule
\textit{\textbf{Language}} & \textit{\textbf{\# Types of Vul}} & \textit{\textbf{\# Projects}} & \textit{\textbf{\# Commits}} & \textit{\textbf{\# Files}} \\
\midrule
C++      & 105      & 169     & 689     & 1,506  \\
Java     & 129      & 307     & 888     & 2,925  \\
Python   & 159      & 366     & 1,310    & 2,760 \\
\bottomrule
\end{tabular}
}
\label{dataset}
\end{table}

\subsubsection{Component}
We provide the details of each component's role and functionality in this section.

\textbf{1. Vulnerability Expert Agents.} 
Vulnerability expert agents for evaluators constitute a critical component in \tool, where each individual LLM is regarded as an expert agent for SV assessment tasks.
For each task related to SV assessment, we meticulously craft unique prompts tailored to the specific requirements of the task.
Each LLM is tasked with evaluating the severity of a vulnerability-related commit and subsequently providing a detailed explanation. 
The responses generated by all agents are preserved within the chat history. 
This archive of interactions enables subsequent evaluators in future rounds of assessment from prior communications, which mirrors the real-world interactions for SV assessment.
It is worth mentioning that each agent evaluates all aspects of the same commit, employing different prompts tailored to specific tasks.

\textbf{2. Communication Strategy.}
Another pivotal challenge involves leveraging references from previous expert analyses to construct new prompts that facilitate further exploration by agents. As previously discussed, assessing the multifaceted aspects of vulnerabilities is an intricate and critical process, we are more concerned with how to refer to other expert responses and interpretations for further SV assessment. As shown in Figure~\ref{strategy}, we explore four distinct communication strategies to emulate the processes in the real-world scenarios for SV assessment.

\textit{(1) Referencing the preceding one expert.} Each expert agent constructs its response based on the input from the immediately preceding expert, except the initial agent. We incorporate only the prior agent's response into the current agent’s conversational history. It prevents excessive past interactions from influencing present SV assessment results.

\textit{(2) Referencing the previous communication.}
The expert agents sequentially generate their responses in a predetermined order. This procedure involves concatenating all previous responses into the chat history to construct the assistant's prompt for the next agents. This approach simulates the written communication for SV assessment in the real world, where experts access all prior information and make their judgments accordingly.

\textit{(3) Simultaneous assessment.}
Every expert agent cannot reference the responses of other experts from the current round but may consider the responses from all experts in the previous round. This method minimizes the dependency of an agent on the responses of other experts and mitigates the influences that could arise from sequential order.

\textit{(4) Summarizer assessment.}
Building on the strategy (3), each round additionally augments a summarizer, which synthesizes the responses of all experts within the current round and makes a final judgment. This approach emulates real-world scenarios where conflicting opinions on SV assessments, and introduces an expert specifically designated for decision-making purposes.

\textbf{3. Adaptive Environment.}
In \tool, 
each LLM is treated as an agent that interacts with the adaptive environment. The environment presents two aspects: integration of knowledge from the CVSS standard and coordination with multiple agents from the chat history.
The CVSS standard, which can be either predefined or user-modified, is designed to facilitate the rapid integration of new domain knowledge and adapt to evolving standards in SV assessment. 
The chat history is dynamically produced by each agent. The responses generated by different agents collaboratively contribute to updates in the prompt, enhancing the collaborative process.

\section{Experiments}

We construct a new benchmark and evaluate the benefits of \tool, intending to understand the following questions:

\textbf{Q1:} How does multi-agent of \tool compare to the single-agent for SV assessment?

\textbf{Q2:} How does \tool perform on different communication strategies for SV assessment?

\textbf{Q3:} What are the impacts of expert numbers and communication rounds in the \tool?

\subsection{Data Preparation}
\label{datapreparation}
Securing high-quality datasets comprising vulnerability-related commits for SV assessment is a formidable challenge, necessitating the demand for qualified expertise. 

\textbf{Data Collection:} Our initial step involved acquiring open-source vulnerabilities from Mend~\cite{Mend}, which provides extensive vulnerability entries contributed by a community of experts. For each identified vulnerability entry, we extracted security-related commits (i.e., patches) from platforms such as GitHub, Android, and Chrome, recording their associated project and commit messages.

\textbf{Data Filter:} 
To ensure the relevance and accuracy of our dataset, we employed a filtering methodology to select commits based on two essential criteria: (1) All SV assessment labels must be complete, and (2) The labels for SV assessments must conform to the evaluation standards established by CVSS V3.1. Additionally, we utilized time-based splits for testing the \tool, aiming to closely mimic real-world scenarios where future unseen data is not available. 

As presented in Table~\ref{dataset}, we have gathered 699, 888, and 1,310 vulnerability-related commits in C++, Python, and Java, respectively. They are collected according to the CVSS v3.1 standard and encompass  105, 129, and 159 types of vulnerabilities across the 160, 307, and 366 projects, respectively.

\subsection{Dataset Evaluation}
\begin{table}[t]
\centering
\setlength{\tabcolsep}{5mm}
\renewcommand{\arraystretch}{1.05}

\caption{Dataset evaluation in C++, Java and Python. \wxc{}}
\resizebox{0.4\textwidth}{!}{
\begin{tabular}{cccc}
\toprule
\textit{\textbf{Datasets}}             & \textit{\textbf{Language}} & \textit{\textbf{Accuracy}} \\
\toprule
Big-Vul~\cite{DBLP:conf/msr/FanL0N20/bigvul}                                & C/C++                      & 54.3                                             \\
D2A~\cite{DBLP:conf/icse/ZhengPLBEYLMS21}                                 & C/C++                      & 28.6                                             \\
\midrule
\multirow{3}{*}{\tool} & C++                       &      90.0                                    \\
                                      & Python                     &   65.0                             \\
                                      & Java                       &                   70.0         \\               
\bottomrule
\end{tabular}
}
\label{humaneval}
\end{table}

The previous study~\cite{DBLP:conf/icse/CroftBK23} has demonstrated that vulnerability datasets often exhibit quality problems. Therefore, we conducted an evaluation of our dataset in comparison with existing datasets, despite the absence of specific datasets dedicated to vulnerability assessment. Specifically, we randomly select 20 examples from each programming language and manually analyze the vulnerability. The manual analysis is independently carried out by two developers, each possessing over five years of experience in software security. As presented in Table~\ref{humaneval}, our dataset demonstrates a higher accuracy compared to previous datasets, underscoring the effectiveness of our data collection and filtration processes.

\subsection{Baselines} 

We primarily focus on few-shot-based methods for SV assessment. 
This is due to the insufficiency of labeled data for CVSS v3.1 available in programming languages such as C++, Java, and Python. Despite the limited data, these languages pose significant vulnerability threats. 

We use the Yin et al.~\cite{DBLP:journals/corr/abs-2404-02056/llmsva} method as baseline, which directly involves a single LLM to generate a response for the given commit (i.e., \baseline). This approach tests the LLM's ability for SV assessment. For the LLMs utilized in \tool, we have selected ChatGPT~\cite{ChatGPT} and GPT-4~\cite{GPT4}, given their robust capabilities in handling code-related tasks. 
 

\subsection{Evaluation Metrics}
In this paper, we employ the evaluation framework delineated by the CVSS v3.1 for SV assessment results derived from various methods. Specifically, we compute the Accuracy (i.e., Acc), which quantifies the ratio of accurately classified instances to the total number of instances, and calculate the F1 score (i.e., F1) to evaluate issues of class imbalance situation. 

\subsection{\tool Results}
\begin{table*}[t]
\centering
\setlength{\tabcolsep}{3.2mm}
\renewcommand{\arraystretch}{1.05}
\resizebox{\textwidth}{!}{
\begin{tabular}{lllcccccccccccccccc}
\toprule
\multicolumn{3}{l}{\textit{\textbf{Exploitability Metrics}}}                                       & \multicolumn{2}{c}{\textbf{Attack Vector}} & \multicolumn{2}{c}{\textbf{Access Complexity}} & \multicolumn{2}{c}{\textbf{Privileges Required}} & \multicolumn{2}{c}{\textbf{User Interaction}} & \multicolumn{2}{c}{\textbf{Average}}\\

\textit{\textbf{Lang}}      & \multicolumn{2}{c}{\textit{\textbf{Baselines}}}    & \textbf{Acc}      & \textbf{F1}      & \textbf{Acc}        & \textbf{F1}        & \textbf{Acc}         & \textbf{F1}         & \textbf{Acc}        & \textbf{F1} & \textbf{Acc}        & \textbf{F1}      \\
\toprule
\multirow{4}{*}{Java}   & \multirow{2}{*}{ChatGPT} & \baseline               & 0.4778               & 0.4075               & 0.4000               & 0.3132               & 0.2889               & 0.2425               & 0.3667               & 0.3532               & 0.3834          & 0.3291          \\
                        &                          & \tool                & 0.4889               & 0.4564               & 0.4444               & 0.2167               & 0.5778               & 0.3613               & 0.5000               & 0.4994               & 0.5028          & 0.3835          \\
                        & \multirow{2}{*}{GPT-4}    & \baseline               & 0.8111               & 0.6052               & 0.5222               & 0.2338               & 0.6444               & 0.4633               & 0.7111               & 0.6928               & 0.6722          & 0.4988          \\
                        &                          & \tool                & \textbf{0.8667}      & \textbf{0.6296}      & \textbf{0.5556}      & \textbf{0.3189}      & \textbf{0.8333}      & \textbf{0.6671}      & \textbf{0.7333}      & \textbf{0.7091}      & \textbf{0.7472} & \textbf{0.5812} \\
\hdashline
\multirow{4}{*}{Python} & \multirow{2}{*}{ChatGPT} & \baseline               & 0.3206               & 0.1909               & 0.2137               & 0.1318               & 0.3359               & 0.3004               & 0.3282               & 0.2984               & 0.2996          & 0.2304          \\
                        &                          & \tool                & 0.3282               & 0.2014               & 0.4351               & 0.2510               & 0.5954               & 0.4761               & 0.4504               & 0.4496               & 0.4523          & 0.3445          \\
                        & \multirow{2}{*}{GPT-4}    & \baseline               & 0.8168               & \textbf{0.3905}      & 0.1985               & 0.1311               & 0.6870               & 0.5516               & 0.6718               & 0.6480               & 0.5935          & 0.4303          \\
                        &                          & \tool                & \textbf{0.8931}      & 0.3656               & \textbf{0.5573}      & \textbf{0.3251}      & \textbf{0.7176}      & \textbf{0.6089}      & \textbf{0.7557}      & \textbf{0.7292}      & \textbf{0.7309} & \textbf{0.5072} \\
\hdashline
\multirow{4}{*}{C++}    & \multirow{2}{*}{ChatGPT} & \baseline               & 0.3333               & 0.3088               & 0.2754               & 0.1873               & 0.1449               & 0.0921               & 0.5072               & 0.4569               & 0.3152          & 0.2613          \\
                        &                          & \tool                & 0.4203               & 0.3611               & 0.4928               & 0.2686               & 0.6957               & 0.3058               & 0.5652               & 0.5629               & 0.5435          & 0.3746          \\
                        & \multirow{2}{*}{GPT-4}    & \baseline               & 0.7971               & 0.5907               & 0.2174               & 0.1970               & 0.8551               & 0.3073               & 0.5652               & 0.5492               & 0.6087          & 0.4111          \\
                        &                          & \tool                & \textbf{0.8551}      & \textbf{0.6025}      & \textbf{0.6667}      & \textbf{0.4705}      & \textbf{0.9420}      & \textbf{0.4851}      & \textbf{0.5942}      & \textbf{0.5741}      & \textbf{0.7645} & \textbf{0.5331} 
   \\
\bottomrule
\specialrule{0em}{2pt}{2pt}
\toprule
\multicolumn{3}{l}{\textit{\textbf{Scope and Impact Metrics}}}                                       & \multicolumn{2}{c}{\textbf{Scope}}          & \multicolumn{2}{c}{\textbf{Confidentiality }}    & \multicolumn{2}{c}{\textbf{Integrity }}            & \multicolumn{2}{c}{\textbf{Availability}}    & \multicolumn{2}{c}{\textbf{Average}} \\ 

\textit{\textbf{Lang}}       & \multicolumn{2}{c}{\textit{\textbf{Baselines}}}     & \textbf{Acc}      & \textbf{F1}      & \textbf{Acc}        & \textbf{F1}        & \textbf{Acc}         & \textbf{F1}         & \textbf{Acc}        & \textbf{F1}   & \textbf{Acc}        & \textbf{F1}    \\
\toprule
\multirow{4}{*}{Java}   & \multirow{2}{*}{ChatGPT} & \baseline               & 0.1444               & 0.1392               & 0.5111               & 0.2591               & 0.4556               & 0.2406               & 0.4556               & 0.2427               & 0.3917          & 0.2204          \\
                        &                          & \tool                & 0.4000               & 0.3619               & 0.5333               & 0.3572               & 0.4889               & 0.2729               & 0.4667               & 0.2390               & 0.4722          & 0.3078          \\
                        & \multirow{2}{*}{GPT-4}    & \baseline               & 0.4778               & 0.4173               & 0.6222               & 0.4305               & 0.5333               & 0.3158               & 0.5667               & 0.3579               & 0.5500          & 0.3804          \\
                        &                          & \tool                & \textbf{0.5556}      & \textbf{0.4591}      & \textbf{0.7222}      & \textbf{0.6458}      & \textbf{0.6556}      & \textbf{0.5261}      & \textbf{0.5667}      & \textbf{0.4153}      & \textbf{0.6250} & \textbf{0.5116} \\
                        \hdashline
\multirow{4}{*}{Python} & \multirow{2}{*}{ChatGPT} & \baseline               & 0.2443               & 0.2133               & 0.5038               & 0.3379               & 0.4504               & 0.3424               & 0.3511               & 0.2471               & 0.3874          & 0.2852          \\
                        &                          & \tool                & 0.4504               & 0.4386               & 0.5115               & 0.4151               & 0.4733               & 0.3914               & 0.4580               & 0.3432               & 0.4733          & 0.3971          \\
                        & \multirow{2}{*}{GPT-4}    & \baseline               & 0.5878               & 0.5326               & 0.6412               & 0.5843               & 0.5191               & 0.4082               & 0.5802               & 0.4328               & 0.5821          & 0.4895          \\
                        &                          & \tool                & \textbf{0.6742}      & \textbf{0.5416}      & \textbf{0.6718}      & \textbf{0.6483}      & \textbf{0.5496}      & \textbf{0.4981}      & \textbf{0.7176}      & \textbf{0.4997}      & \textbf{0.6533} & \textbf{0.5469} \\
                        \hdashline
\multirow{4}{*}{C++}    & \multirow{2}{*}{ChatGPT} & \baseline               & 0.1449               & 0.1384               & 0.4638               & 0.2622               & 0.4928               & 0.3192               & 0.5072               & 0.2509               & 0.4022          & 0.2427          \\
                        &                          & \tool                & 0.5217               & 0.3907               & 0.5507               & 0.4029               & 0.5217               & 0.3436               & 0.6812               & 0.4059               & 0.5688          & 0.3858          \\
                        & \multirow{2}{*}{GPT-4}    & \baseline               & 0.6087               & 0.3784               & \textbf{0.6087}      & 0.4230               & 0.6087               & 0.3739               & 0.7101               & 0.3778               & 0.6341          & 0.3883          \\
                        &                          & \tool                & \textbf{0.7246}      & \textbf{0.5043}      & \textbf{0.6087}      & \textbf{0.5163}      & \textbf{0.6812}      & \textbf{0.5935}      & \textbf{0.7246}      & \textbf{0.4825}      & \textbf{0.6848} & \textbf{0.5242}
\\

\bottomrule
\end{tabular}}
\caption{
Experimental results of \tool across Java, Python and C++. We \textbf{bold} the best-performing method for each metric.
}
\label{RQ1}
\end{table*}

As illustrated in Table~\ref{RQ1}, these LLM-based approach tasks are to achieve consistency with the SV assessment results of human experts in the CVSS v3.1 framework. Our findings reveal that:
\textbf{(1) SV assessment is an arduous task for a single agent}. Existing single-based LLMs perform poorly across all metrics SV assessment with commit input, with average performances as low as 48.50\% and 34.73\% on the accuracy and F1 metrics, respectively.
This underscores the complexity and difficulty of SV assessment for the single LLM.
\textbf{(2) Superior performance of \tool.} \tool significantly enhances the performance of the SV assessment process, achieving higher alignment with human preference compared to single-agent-based methods. Specifically, the multi-agent-based method improves the F1 by 53.71\% for ChatGPT and 32.88\% for GPT-4. This demonstrates \tool's advanced ability to evaluate the different aspects of SV assessment.
\textbf{(3) GPT-4 can aid human experts in SV assessment.} The ChatGPT method shows more substantial improvements in the exploitability aspect, with average increases of 72.35\% and 49.35\%, respectively. In contrast, the GPT-4 shows more significant improvements on the impact metric, with an absolute improvement of 5.64\% and 12.64\% on the accuracy and F1 Score, respectively. Overall, GPT-4 performs well on the \textit{AV}, \textit{PR}, and \textit{UI} metrics, significantly aiding human experts in SV assessment.
\textbf{(4) SV assessment in Python and Java presents the greatest challenge.} Language-specific results reveal that C++ tasks typically exhibit higher accuracy than Python and Java across all multi-agent methods. This discrepancy may be attributed to C++ providing features like manual memory management and extensive use of pointers.
However, these same complexities might make it easier for \tool to SV assessment because they follow certain patterns typical to C++ programming.


We also study the different types of vulnerabilities misreported by \tool.
Despite achieving optimal results in various scenarios, we find that \tool still exhibits an error rate with certain types of vulnerabilities, notably those related to XML. 
For instance, 
\tool incorrectly reported seven instances of CWE-79~\cite{CWE-79} vulnerabilities in Python, and the single agent reported 23 false positives showing a more severe error rate.
This count is the highest among all types of vulnerabilities misreported in terms of the AC metric. 
Furthermore, both single agent and \tool record 15 false positives in the confidentiality metric, which underscores the ongoing need for \tool to enhance its detection capabilities for XML vulnerabilities.

These findings suggest the potential benefits of incorporating corresponding-related code snippets for expert agents to better assess language-specific vulnerabilities, particularly for programming languages with complex structures like Python and Java.

\begin{table*}[t]
\centering
\setlength{\tabcolsep}{4mm}
\renewcommand{\arraystretch}{1.08}
\resizebox{\textwidth}{!}{
\begin{tabular}{lcccccccccc}
\toprule
{\textit{\textbf{Exploitability Metrics}}}                                       & \multicolumn{2}{c}{\textbf{Attack Vector}} & \multicolumn{2}{c}{\textbf{Access Complexity}} & \multicolumn{2}{c}{\textbf{Privileges Required}} & \multicolumn{2}{c}{\textbf{User Interaction}} & \multicolumn{2}{c}{\textbf{Average}}       \\

\textit{\textbf{Communication Strategy}} & \textbf{Acc} & \textbf{F1} & \textbf{Acc} & \textbf{F1} & \textbf{Acc} & \textbf{F1} & \textbf{Acc} & \textbf{F1} & \textbf{Acc} & \textbf{F1}\\
\toprule
Single Agent            & 0.3206          & 0.1909          & 0.2137          & 0.1318          & 0.3359          & 0.3004          & 0.3282          & 0.2984          & 0.2996          & 0.2304          \\
Previous Communication  & 0.2366          & 0.1509          & \textbf{0.4351} & \textbf{0.2510} & \textbf{0.5954} & \textbf{0.4761} & 0.3511          & 0.3487          & 0.4046          & 0.3067          \\
Preceding One Expert    & 0.2137          & 0.1398          & 0.3893          & 0.2181          & 0.5496          & 0.4607          & \textbf{0.4504} & 0.4465          & 0.4008          & 0.3163          \\
Simultaneous Assessment & 0.2672          & 0.1697          & 0.4580          & 0.2494          & 0.5878          & 0.4710          & 0.4427          & 0.4426          & \textbf{0.4389} & 0.3332          \\
Summarizer Assessment   & \textbf{0.3282} & \textbf{0.2014} & 0.4122          & 0.2310          & 0.5573          & 0.4552          & \textbf{0.4504} & \textbf{0.4496} & 0.4370          & \textbf{0.3343} \\
\bottomrule
\specialrule{0em}{2pt}{2pt}
\toprule
\multicolumn{1}{l}{\textit{\textbf{Scope and Impact Metrics}}} & \multicolumn{2}{c}{\textbf{Scope}}          & \multicolumn{2}{c}{\textbf{Confidentiality }}    & \multicolumn{2}{c}{\textbf{Integrity}}            & \multicolumn{2}{c}{\textbf{Availability}}    & \multicolumn{2}{c}{\textbf{Average}}\\

\textit{\textbf{Communication Strategy}} & \textbf{Acc} & \textbf{F1} & \textbf{Acc} & \textbf{F1} & \textbf{Acc} & \textbf{F1} & \textbf{Acc} & \textbf{F1}    & \textbf{Acc} & \textbf{F1}     \\
\toprule
Single Agent            & 0.2443          & 0.2133          & 0.5038          & 0.3379          & 0.4504          & 0.3424          & 0.3511          & 0.2471          & 0.3874          & 0.2852          \\
Previous Communication  & 0.4198          & 0.4190          & 0.4656          & 0.3624          & 0.4275          & 0.3039          & 0.4351          & 0.2921          & 0.4370          & 0.3444          \\
Preceding One Expert    & \textbf{0.4504} & \textbf{0.4386} & \textbf{0.5115} & \textbf{0.4151} & \textbf{0.4733} & \textbf{0.3914} & \textbf{0.4580} & \textbf{0.3432} & \textbf{0.4733} & \textbf{0.3971} \\
Simultaneous Assessment & 0.4351          & 0.4311          & 0.4733          & 0.3445          & 0.4122          & 0.2879          & 0.4351          & 0.3102          & 0.4389          & 0.3434          \\
Summarizer Assessment   & \textbf{0.4504} & 0.4329          & 0.4122          & 0.2949          & 0.4198          & 0.3368          & 0.4122          & 0.2645          & 0.4237          & 0.3323      
\\  

\bottomrule
\end{tabular}}
\caption{
Experimental results of different communication strategies of ChatGPT in Python. 
We bold the best-performing communication strategy for each metric.
}
\label{RQ2}
\end{table*}

\subsection{Communication Strategy}
To answer Q2, we propose four different communication strategies termed as \textit{preceding one expert}, \textit{previous communication}, \textit{simultaneous assessment}, and \textit{summarizer assessment} for the SV assessment task.
We experiment with these strategies in Python and the detailed results are described in Table~\ref{RQ2}. 
The remaining experiment results of Java and C++ are presented in Appendix.
Our observations indicate that
\textbf{(1) Employing either communication strategy proves advantageous for SV assessment.}
Integrating a multi-agent strategy with ChatGPT results in an improvement of 8.83\% and 8.07\% in accuracy and F1 score, demonstrating the effectiveness of the communication strategy methodology, respectively.
\textbf{(2) The efficacy of distinct communication strategies should be tailored to the tasks.} 
Communication strategies exhibit varying performance depending on the task configuration,  which can be attributed to the inherent nature of these tasks.
For instance, the evaluation of \textit{attack complexity} and \textit{user interaction} typically falls under binary classification,
whereas the impact aspect (including \textit{confidentiality}, \textit{integrity}, and \textit{availability}) requires multi-classification.
This underscores the necessity of adopting task-specific communication strategies in the development of SV assessment methods. 
\textbf{(3) The superior performance of \textit{preceding one expert} strategy for most metrics.} \textit{Preceding one expert} strategy demonstrates superior performance in four tasks, yielding significant F1 improvements of 1.32\%, 14.54\%, 14.31\%, and 10.64\% in \textit{scope}, \textit{confidentiality}, \textit{integrity}, and \textit{availability}, respectively. It suggests that excessive reliance on the previous references may lead to deviations in the understanding of expert agents. 
\subsection{Expert Numbers and Communication Rounds}
To answer Q3, we conduct the experiment to study the influence of different expert numbers and communication rounds for assessing vulnerability. 

\definecolor{myblue}{RGB}{20, 80, 170}
\definecolor{mypurple}{RGB}{148, 103, 189}
\definecolor{myorange}{RGB}{255, 127, 14}
\definecolor{mygreen}{RGB}{44, 190, 44}
\begin{figure}
\hspace{-1mm}
\begin{minipage}{0.23\textwidth}
    \centering
    \begin{tikzpicture}
      \begin{axis}[
        width=\linewidth,
        xtick={1,2,3,4,5},
        xticklabels={1,2,3,4,5},
        ytick={0.3, 0.4, 0.5}
      ]
\addplot[mark=o, myblue, thick] coordinates {(1,0.3556) (2,0.2889) (3,0.3333) (4,0.2333) (5,0.2556)};
\addplot[mark=square, mypurple, thick] coordinates {(1,0.4931) (2,0.5431) (3,0.5833) (4,0.5741) (5,0.533)};
\addplot[mark=triangle*, myorange, thick] coordinates {(1,0.4872) (2,0.5545) (3,0.6154) (4,0.5577) (5,0.5353)};
\addplot[mark=star, mygreen, thick] coordinates {(1,0.3345) (2,0.2875) (3,0.3304) (4,0.2325) (5,0.2555)};
      \end{axis}
    \end{tikzpicture}
    \label{hyper:1}
    \subcaption{Expert number of AV.}
  \end{minipage}
  \hspace{-1mm}
  \begin{minipage}{0.23\textwidth}
    \centering
    \begin{tikzpicture}
      \begin{axis}[
        width=\linewidth,
        xtick={1,2,3,4,5},
        xticklabels={1,2,3,4,5},
        ytick={0.3, 0.4, 0.5}
      ]

\addplot[mark=o, myblue, thick] coordinates {(1,0.5556) (2,0.5667) (3,0.5444
) (4,0.4778) (5,0.4778)};

\addplot[mark=square, mypurple, thick] coordinates {(1,0.2832) (2,0.4275) (3,0.3569) (4,0.3250) (5,0.362)};

\addplot[mark=triangle*, myorange, thick] coordinates {(1,0.2751) (2,0.5074) (3,0.4596
) (4,0.4195) (5,0.4476)};

\addplot[mark=star, mygreen, thick] coordinates {(1,0.2758) (2,0.4063) (3,0.3600) (4,0.3128 ) (5,0.3397)};
      \end{axis}
    \end{tikzpicture}
    \label{hyper:2}
    \subcaption{Expert number of PR.}
  \end{minipage}



\hspace{-1mm}
\begin{minipage}{0.23\textwidth}
    \centering
    \begin{tikzpicture}
      \begin{axis}[
        width=\linewidth,
        xtick={1,2,3,4,5},
        xticklabels={1,2,3,4,5},
        ytick={0.3, 0.4, 0.5}
      ]
\addplot[mark=o, myblue, thick] coordinates {(1,0.3222) (2,0.2889) (3,0.2889) (4,0.3111) (5,0.2667)};

\addplot[mark=square, mypurple, thick] coordinates {(1,0.5511) (2,0.5431) (3,0.5431) (4,0.5214) (5,0.5367)};

\addplot[mark=triangle*, myorange, thick] coordinates {(1,0.5737) (2,0.5545) (3,0.5545) (4,0.5321) (5,0.5417)};

\addplot[mark=star, mygreen, thick] coordinates {(1,0.3181) (2,0.2875) (3,0.2875) (4,0.3056) (5,0.2663)};
      \end{axis}
    \end{tikzpicture}
    \label{hyper:4}
    \subcaption{Round of AV.}
  \end{minipage}
  \hspace{-1mm}
  \begin{minipage}{0.23\textwidth}
    \centering
    \begin{tikzpicture}
      \begin{axis}[
        width=\linewidth,
        xtick={1,2,3,4,5},
        xticklabels={1,2,3,4,5},
        ytick={0.3, 0.4, 0.5}
      ]
\addplot[mark=o, myblue, thick] coordinates {(1,0.5333) (2,0.5444) (3,0.5889) (4,0.5778) (5,0.5)};

\addplot[mark=square, mypurple, thick] coordinates {(1,0.3553) (2,0.4045) (3,0.4121) (4,0.4154) (5,0.37)};

\addplot[mark=triangle*, myorange, thick] coordinates {(1,0.3023) (2,0.469) (3,0.5082) (4,0.4938) (5,0.2963)};

\addplot[mark=star, mygreen, thick] coordinates {(1,0.3266) (2,0.3689) (3,0.4093) (4,0.3941) (5,0.3288)};

      \end{axis}
    \end{tikzpicture}
    \label{hyper:5}
    \subcaption{Round of PR.}
  \end{minipage}



\caption{The impact of expert numbers and communication rounds on \tool in the Java dataset. The \textcolor{myblue}{blue}, \textcolor{mypurple}{purple}, \textcolor{myorange}{orange}, and \textcolor{mygreen}{green} lines denote the \textcolor{myblue}{accuracy}, \textcolor{mypurple}{precision}, \textcolor{myorange}{recall} and \textcolor{mygreen}{F1 score} metrics (higher is better), respectively.}
\label{RQ3}
\end{figure}
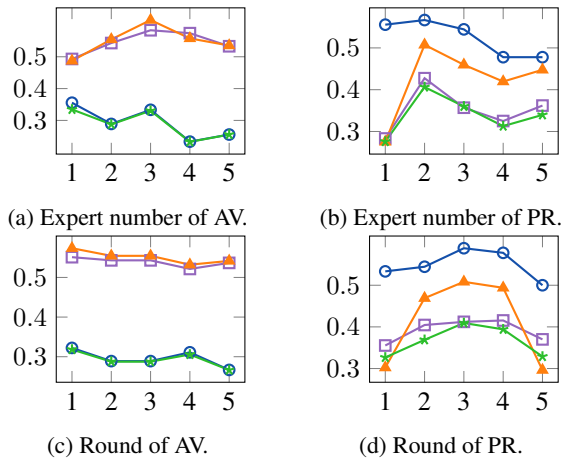
\textbf{Expert Numbers.}
The number of experts should be selected as medium (2-3). As illustrated in Figure~\ref{RQ3} (a)-(b), the correlation between the number of experts and performance demonstrates a pattern of initial improvement followed by a subsequent decrease, with the optimal performance occurring when the number of experts is 2-3.
This suggests that diverse expert roles enhance the model's comprehension of SV assessments, aligning with findings reported by~\cite{DBLP:journals/corr/abs-2305-14325, DBLP:journals/corr/abs-2308-07201/Chateval}. Furthermore, it indicates that an excessive number of experts involved in the decision-making process may misguide the LLM-based method decisions, potentially due to the extended context length.

\textbf{Communication Rounds.}
Multiple rounds of communication are required to facilitate the model's understanding of vulnerability due to its lack of domain-specific knowledge. However, communication across numerous rounds does not necessarily even result in a decline. This could be attributed to the fact that excessively long contexts are detrimental to the model's ability to effectively process the task of SV assessment. It is noteworthy that different tasks may necessitate varying numbers of communication rounds. For instance, the \textit{PR} exhibits optimal performance after three rounds, while \textit{AV} reaches peak performance in the first rounds. These findings underscore the need for a more sophisticated appreciation of the balance between the number of communication rounds and the specific task to optimize performance.

\section{Discussion}
\subsection{Case Study}
\begin{figure}[t]
	\centering
    \includegraphics[width=0.5\textwidth]{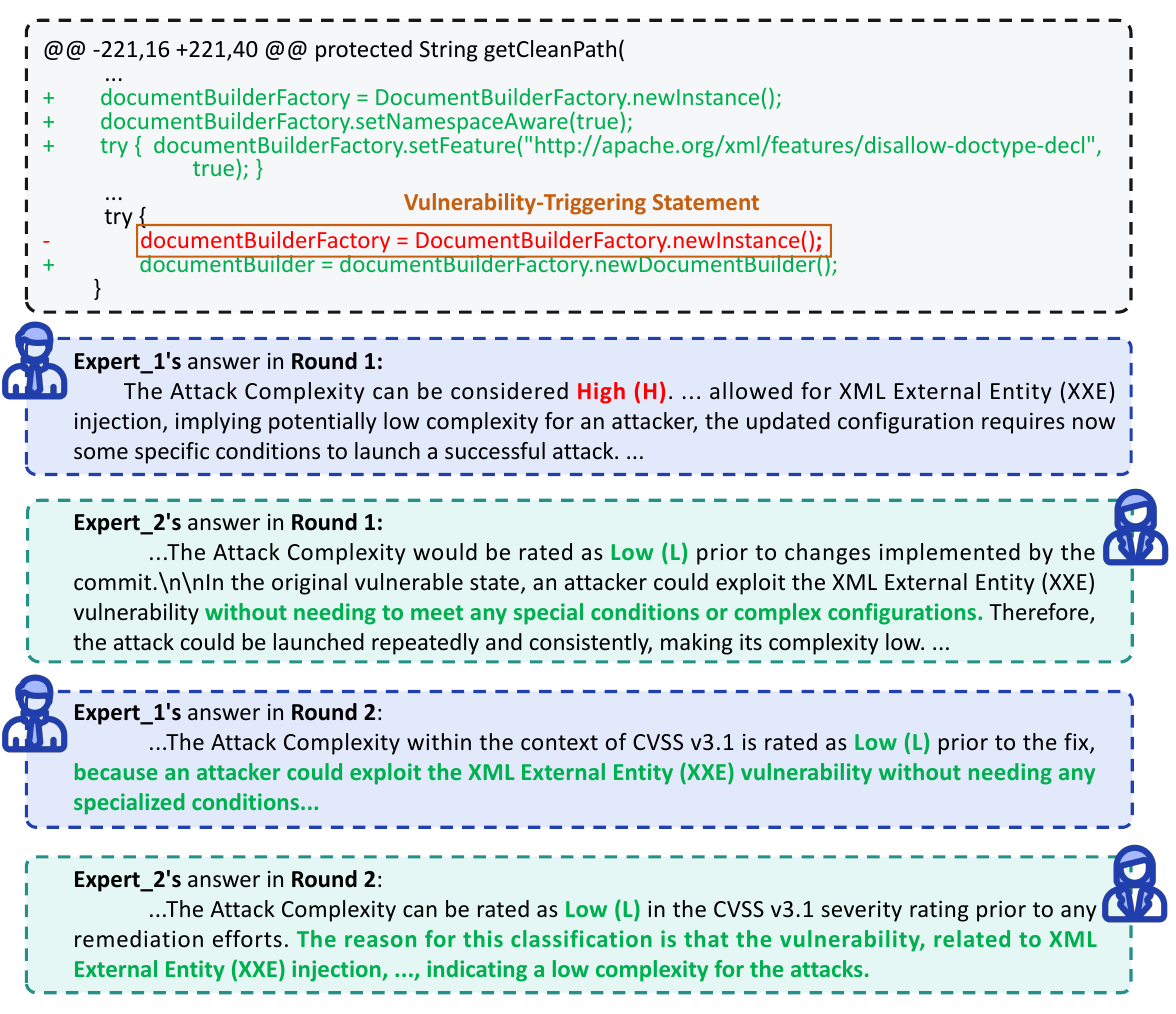}
	\caption{A \tool example presents a discussion process involving two expert agents. The text shaded in \textcolor{red}{red} and \textcolor{darkgreen}{green} denote the \textcolor{red}{wrong} and \textcolor{darkgreen}{right} responses from LLMs, respectively. 
 }
	\label{casestudy}
\end{figure}

Figure~\ref{casestudy} is a vulnerability example from CVE-2023-46502~\cite{CVE-2023-46502}, which uses the \tool to evaluate the \textit{Attack Complexity}.
The vulnerability arises from the improper configuration of \textit{DocumentBuilderFactory} (shaded in brown in Figure~\ref{casestudy}), which allows XML external entity attacks (i.e., CWE-611~\cite{CWE-611}).
We observe that there initially exists a discrepancy in opinions between the different agents during the first round of responses. Then, a consensus is reached in the subsequent round. This case mirrors real-world situations where multiple experts assess a single vulnerability.
Specifically, \tool demonstrates several human-like decision-making processes observed in the industry. \textbf{(1) Opinions diversity}: Initially, Expert 1 and Expert 2 present differing judgments when assessing the same vulnerability commit. This diversity broadens the perspective and encompasses a more comprehensive range of considerations in SV assessment.
\textbf{(2) Revision}: 
Upon considering the viewpoints of other experts, Expert 1 learns from different aspects and revises its previously erroneous judgment. This indicates that \tool, when informed by the perspectives of multiple experts, possesses the capability to revise.
\textbf{(3) Interpretability}: Each expert provides explanations for their assessments. This practice aligns with industry standards set by FIRST~\cite{first}, which mandates that CVSS must adhere to documented guidelines and include both the scoring vector and a detailed rationale, enabling others to understand the derivation of the scores. Previous methods~\cite{DBLP:conf/kbse/LeHCB21/DeepCVA, DBLP:conf/sigsoft/LiYZWN23} often provided scores without the explanations needed for comprehensive SV assessment.
\textbf{(4) Evolutionary adaptation}: \tool can be adapted to different versions of SV assessment systems based on the prompts.
Unlike prior works, \tool swiftly integrates current version-specific domain knowledge to conduct SV assessments without training, demonstrating its agility and relevance in evolving systems.

\subsection{Limitation}
\label{limitations}
\textit{Transferability on other types of SV assessment.} 
In this paper, we only focus on SV assessment with commit input and CVSS v3.1 standard, excluding SV and bug report-based methods. However, \tool can be generalized to other types of SV assessment and other expert-based SV assessment standards.
In the future, we intend to explore the efficacy of \tool regarding the upgrade of the assessment system.

\textit{Constraints of domain knowledge in prompts.}
For the context limited of LLMs, \tool only contains the prompt-based domain knowledge and chat history to facilitate the SV assessment. In the future, we will explore more expert-based examples as prompts for the LLM-based SV assessment.

\section{Related Work}
Public security databases, such as the NVD~\cite{NVD}, and expert-based scoring systems, such as the CVSS~\cite{CVSSwhat} have been pivotal in furnishing detailed datasets for SV. In recent years, the CVSS framework has witnessed significant enhancements~\cite{DBLP:conf/ic-nc/FeutrillRYR18}, evolving from v2~\cite{CVSSmetricsV2} to v3.0~\cite{CVSSmetricsV3}, and subsequently to v3.1~\cite{CVSSmetrics}. 
Specifically, the existing methods can be broadly divided into two aspects: SV report-based and commit-based methods. 
The majority of existing methods for automated SV assessment depend on SV reports (i.e., SV reported-based methods)~\cite{DBLP:conf/icsm/HanLXLF17, DBLP:conf/msr/LamkanfiDGG10, DBLP:conf/msr/LeSB19, DBLP:journals/jss/SpanosA18} from the NVD. 
These methods typically focus on predicting either a single metric~\cite{DBLP:journals/ese/FuTLKNPG24, DBLP:conf/qrs/KudjoCZMH19, DBLP:conf/icaci/WangZSZ19} or a set of metrics associated with the CVSS~\cite{DBLP:conf/msr/LeB22, DBLP:conf/badgers/YamamotoMN15, DBLP:conf/kbse/OgnawalaAPK18}. For instance, Han et al.~\cite{DBLP:conf/icsm/HanLXLF17} introduced a Convolutional Neural Network-based method to automate and predict the overall severity rating by analyzing SV descriptions.  
However, these user-submitted SV reports often exhibit significant delays~\cite{DBLP:conf/icsm/ThungLJLRD12, DBLP:journals/corr/abs-2112-10123, DBLP:conf/oopsla/BosuC12, DBLP:conf/msr/ThongtanunamMHI15}, potentially exceeding 1000 days. 
To expedite SV assessment and reduce the extensive labor required by human experts for evaluations, 
In addition, the recent research also explored the potential of commit-based methods~\cite{DBLP:conf/kbse/LeHCB21/DeepCVA, DBLP:journals/tosem/ZhouSWLL22, DBLP:conf/sigsoft/LiYZWN23, DBLP:journals/corr/abs-2404-02056}. This type of method involves utilizing commit changes to assess all aspects of SVs. For instance, Le et al.~\cite{DBLP:conf/kbse/LeHCB21/DeepCVA} introduced DeepCVA, a model that applies multi-task learning to perform commit-based SV assessment. Li et al.~\cite{DBLP:conf/sigsoft/LiYZWN23} proposed a neural framework dedicated to SV detection and assessment simultaneously. 


Considering that SV assessment systems are subject to continuous evolution or require customization tailored to specific application contexts, it is imperative for a practical framework to demonstrate effectiveness in limited labels.  In this paper, we propose a multi-agent framework for SV assessment to tackle the challenge of lacking data labels under the new standards. 
Furthermore, we construct a new dataset that includes CVSS v3.1 assessment metrics across multi-lingual programming languages.


\section{Conclusion}

In this paper, we propose the first multi-agent-based framework \tool to simulate vulnerability assessment strategies in real-world scenarios. Furthermore, we construct the first multi-lingual SV assessment dataset based on the new standard of CVSS, comprising 699, 888, and 1,310 vulnerability-related commits in C++, Python, and Java, respectively, which can serve as a foundation dataset for future research.
We emphasize the necessity of developing multi-agent evaluators for SV assessment due to the continuous evolution of CVSS. Our experimental results confirm the effectiveness of \tool, especially in scenarios with limited labeled data.
We also find that \tool offers a human-like process, providing both rationale and responses for SV assessment. This underscores the effectiveness and possibility of \tool for the next generation of SV assessment.
\bibliography{aaai25}

\begin{thebibliography}{58}
\providecommand{\natexlab}[1]{#1}

\bibitem[{CVS(2023)}]{CVSSwhat}
 2023.
\newblock What is CVSS score.
\newblock \url{https://debricked.com/blog/what-is-cvss-score/}.

\bibitem[{CVS(2024{\natexlab{a}})}]{CVSS}
 2024{\natexlab{a}}.
\newblock Common Vulnerability Scoring System (CVSS).
\newblock \url{https://www.first.org/cvss/}.

\bibitem[{fir(2024)}]{first}
 2024.
\newblock Common Vulnerability Scoring System SIG.
\newblock \url{https://www.first.org/cvss/}.

\bibitem[{CVS(2024{\natexlab{b}})}]{CVSSmetricsV3}
 2024{\natexlab{b}}.
\newblock Common Vulnerability Scoring System v3.0: Specification Document.
\newblock \url{https://www.first.org/cvss/v3.0/specification-document}.

\bibitem[{CVS(2024{\natexlab{c}})}]{CVSSmetrics}
 2024{\natexlab{c}}.
\newblock Common Vulnerability Scoring System v3.1: Specification Document.
\newblock \url{https://www.first.org/cvss/v3.1/specification-document}.

\bibitem[{CVS(2024{\natexlab{d}})}]{CVSSmetricsV2}
 2024{\natexlab{d}}.
\newblock A Complete Guide to the Common Vulnerability Scoring System Version 2.0.
\newblock \url{https://www.first.org/cvss/v2/guide}.

\bibitem[{CWE(2024{\natexlab{a}})}]{CWE-611}
 2024{\natexlab{a}}.
\newblock CWE-611: Improper Restriction of XML External Entity Reference.
\newblock \url{https://cwe.mitre.org/data/definitions/611.html}.

\bibitem[{CWE(2024{\natexlab{b}})}]{CWE-79}
 2024{\natexlab{b}}.
\newblock CWE-79: Improper Neutralization of Input During Web Page Generation ('Cross-site Scripting').
\newblock \url{https://cwe.mitre.org/data/definitions/79.html}.

\bibitem[{CVE(2024)}]{CVE}
 2024.
\newblock “{Common Vulnerabilities and Exposures (CVE)}”.
\newblock \url{https://cve.mitre.org/}.

\bibitem[{Bilge and Dumitras(2012)}]{DBLP:conf/ccs/BilgeD12}
Bilge, L.; and Dumitras, T. 2012.
\newblock Before we knew it: an empirical study of zero-day attacks in the real world.
\newblock In Yu, T.; Danezis, G.; and Gligor, V.~D., eds., \emph{the {ACM} Conference on Computer and Communications Security, CCS'12, Raleigh, NC, USA, October 16-18, 2012}, 833--844. {ACM}.

\bibitem[{Bosu and Carver(2012)}]{DBLP:conf/oopsla/BosuC12}
Bosu, A.; and Carver, J.~C. 2012.
\newblock Peer code review in open source communitiesusing reviewboard.
\newblock In Murphy{-}Hill, E.~R.; Sadowski, C.; and Markstrum, S., eds., \emph{Proceedings of the {ACM} 4th Annual Workshop on Evaluation and Usability of Programming Languages and Tools, {PLATEAU} 2012, Tucson, AZ, USA, October 21, 2012}, 17--24. {ACM}.

\bibitem[{Chan et~al.(2023)Chan, Chen, Su, Yu, Xue, Zhang, Fu, and Liu}]{DBLP:journals/corr/abs-2308-07201/Chateval}
Chan, C.; Chen, W.; Su, Y.; Yu, J.; Xue, W.; Zhang, S.; Fu, J.; and Liu, Z. 2023.
\newblock ChatEval: Towards Better LLM-based Evaluators through Multi-Agent Debate.
\newblock \emph{CoRR}, abs/2308.07201.

\bibitem[{ChatGPT(2022)}]{ChatGPT}
ChatGPT. 2022.
\newblock ChatGPT.
\newblock \url{https://chat.openai.com/}.

\bibitem[{Croft, Babar, and Kholoosi(2023)}]{DBLP:conf/icse/CroftBK23}
Croft, R.; Babar, M.~A.; and Kholoosi, M.~M. 2023.
\newblock Data Quality for Software Vulnerability Datasets.
\newblock In \emph{45th {IEEE/ACM} International Conference on Software Engineering, {ICSE} 2023, Melbourne, Australia, May 14-20, 2023}, 121--133. {IEEE}.

\bibitem[{Croft et~al.(2021)Croft, Newlands, Chen, and Babar}]{DBLP:conf/esem/CroftNCB21}
Croft, R.; Newlands, D.; Chen, Z.; and Babar, M.~A. 2021.
\newblock An Empirical Study of Rule-Based and Learning-Based Approaches for Static Application Security Testing.
\newblock In Lanubile, F.; Kalinowski, M.; and Baldassarre, M.~T., eds., \emph{{ESEM} '21: {ACM} / {IEEE} International Symposium on Empirical Software Engineering and Measurement, Bari, Italy, October 11-15, 2021}, 8:1--8:12. {ACM}.

\bibitem[{Deng et~al.(2024)Deng, Xia, Yang, Zhang, Yang, and Zhang}]{DBLP:conf/icse/DengXYZY024}
Deng, Y.; Xia, C.~S.; Yang, C.; Zhang, S.~D.; Yang, S.; and Zhang, L. 2024.
\newblock Large Language Models are Edge-Case Generators: Crafting Unusual Programs for Fuzzing Deep Learning Libraries.
\newblock In \emph{Proceedings of the 46th {IEEE/ACM} International Conference on Software Engineering, {ICSE} 2024, Lisbon, Portugal, April 14-20, 2024}, 70:1--70:13. {ACM}.

\bibitem[{Detail(2024{\natexlab{a}})}]{CVE-2023-2954}
Detail, C.-.-. 2024{\natexlab{a}}.
\newblock \url{https://nvd.nist.gov/vuln/detail/CVE-2023-2954/}.

\bibitem[{Detail(2024{\natexlab{b}})}]{CVE-2023-46502}
Detail, C.-.-. 2024{\natexlab{b}}.
\newblock \url{https://nvd.nist.gov/vuln/detail/CVE-2023-46502}.

\bibitem[{Dissanayake et~al.(2022)Dissanayake, Jayatilaka, Zahedi, and Babar}]{DBLP:conf/kbse/DissanayakeJZB22}
Dissanayake, N.; Jayatilaka, A.; Zahedi, M.; and Babar, M.~A. 2022.
\newblock An Empirical Study of Automation in Software Security Patch Management.
\newblock In \emph{37th {IEEE/ACM} International Conference on Automated Software Engineering, {ASE} 2022, Rochester, MI, USA, October 10-14, 2022}, 7:1--7:13. {ACM}.

\bibitem[{Du et~al.(2023)Du, Li, Torralba, Tenenbaum, and Mordatch}]{DBLP:journals/corr/abs-2305-14325}
Du, Y.; Li, S.; Torralba, A.; Tenenbaum, J.~B.; and Mordatch, I. 2023.
\newblock Improving Factuality and Reasoning in Language Models through Multiagent Debate.
\newblock \emph{CoRR}, abs/2305.14325.

\bibitem[{Fan et~al.(2020)Fan, Li, Wang, and Nguyen}]{DBLP:conf/msr/FanL0N20/bigvul}
Fan, J.; Li, Y.; Wang, S.; and Nguyen, T.~N. 2020.
\newblock A {C/C++} Code Vulnerability Dataset with Code Changes and {CVE} Summaries.
\newblock In Kim, S.; Gousios, G.; Nadi, S.; and Hejderup, J., eds., \emph{{MSR} '20: 17th International Conference on Mining Software Repositories, Seoul, Republic of Korea, 29-30 June, 2020}, 508--512. {ACM}.

\bibitem[{Feutrill et~al.(2018)Feutrill, Ranathunga, Yarom, and Roughan}]{DBLP:conf/ic-nc/FeutrillRYR18}
Feutrill, A.; Ranathunga, D.; Yarom, Y.; and Roughan, M. 2018.
\newblock The Effect of Common Vulnerability Scoring System Metrics on Vulnerability Exploit Delay.
\newblock In \emph{Sixth International Symposium on Computing and Networking, {CANDAR} 2018, Takayama, Japan, November 23-27, 2018}, 1--10. {IEEE} Computer Society.

\bibitem[{Foreman.(2019)}]{Park}
Foreman., P. 2019.
\newblock Vulnerability management.
\newblock Auerbach Publications.

\bibitem[{Fu et~al.(2024)Fu, Tantithamthavorn, Le, Kume, Nguyen, Phung, and Grundy}]{DBLP:journals/ese/FuTLKNPG24}
Fu, M.; Tantithamthavorn, C.; Le, T.; Kume, Y.; Nguyen, V.; Phung, D.~Q.; and Grundy, J.~C. 2024.
\newblock AIBugHunter: {A} Practical tool for predicting, classifying and repairing software vulnerabilities.
\newblock \emph{Empir. Softw. Eng.}, 29(1): 4.

\bibitem[{Gao et~al.(2023)Gao, Wen, Gao, Wang, Zhang, and Lyu}]{DBLP:conf/kbse/GaoWGWZL23}
Gao, S.; Wen, X.; Gao, C.; Wang, W.; Zhang, H.; and Lyu, M.~R. 2023.
\newblock What Makes Good In-Context Demonstrations for Code Intelligence Tasks with LLMs?
\newblock In \emph{38th {IEEE/ACM} International Conference on Automated Software Engineering, {ASE} 2023, Luxembourg, September 11-15, 2023}, 761--773. {IEEE}.

\bibitem[{Han et~al.(2017)Han, Li, Xing, Liu, and Feng}]{DBLP:conf/icsm/HanLXLF17}
Han, Z.; Li, X.; Xing, Z.; Liu, H.; and Feng, Z. 2017.
\newblock Learning to Predict Severity of Software Vulnerability Using Only Vulnerability Description.
\newblock In \emph{2017 {IEEE} International Conference on Software Maintenance and Evolution, {ICSME} 2017, Shanghai, China, September 17-22, 2017}, 125--136. {IEEE} Computer Society.

\bibitem[{Huang et~al.(2024)Huang, Li, Lam, Liang, Wang, Yuan, Jiao, Wang, Tu, and Lyu}]{DBLP:journals/corr/abs-2403-11807}
Huang, J.; Li, E.~J.; Lam, M.~H.; Liang, T.; Wang, W.; Yuan, Y.; Jiao, W.; Wang, X.; Tu, Z.; and Lyu, M.~R. 2024.
\newblock How Far Are We on the Decision-Making of LLMs? Evaluating LLMs' Gaming Ability in Multi-Agent Environments.
\newblock \emph{CoRR}, abs/2403.11807.

\bibitem[{Imran, Chatterjee, and Damevski(2023)}]{DBLP:journals/corr/abs-2312-09731}
Imran, M.~M.; Chatterjee, P.; and Damevski, K. 2023.
\newblock Uncovering the Causes of Emotions in Software Developer Communication Using Zero-shot LLMs.
\newblock \emph{CoRR}, abs/2312.09731.

\bibitem[{Karpinska, Akoury, and Iyyer(2021)}]{DBLP:conf/emnlp/KarpinskaAI21}
Karpinska, M.; Akoury, N.; and Iyyer, M. 2021.
\newblock The Perils of Using Mechanical Turk to Evaluate Open-Ended Text Generation.
\newblock In Moens, M.; Huang, X.; Specia, L.; and Yih, S.~W., eds., \emph{Proceedings of the 2021 Conference on Empirical Methods in Natural Language Processing, {EMNLP} 2021, Virtual Event / Punta Cana, Dominican Republic, 7-11 November, 2021}, 1265--1285. Association for Computational Linguistics.

\bibitem[{Khan and Parkinson(2018)}]{DBLP:series/ccn/KhanP18}
Khan, S.; and Parkinson, S. 2018.
\newblock Review into State of the Art of Vulnerability Assessment using Artificial Intelligence.
\newblock In Parkinson, S.; Crampton, A.; and Hill, R., eds., \emph{Guide to Vulnerability Analysis for Computer Networks and Systems - An Artificial Intelligence Approach}, Computer Communications and Networks, 3--32. Springer.

\bibitem[{Kudjo et~al.(2019)Kudjo, Chen, Zhou, Mensah, and Huang}]{DBLP:conf/qrs/KudjoCZMH19}
Kudjo, P.~K.; Chen, J.; Zhou, M.; Mensah, S.; and Huang, R. 2019.
\newblock Improving the Accuracy of Vulnerability Report Classification Using Term Frequency-Inverse Gravity Moment.
\newblock In \emph{19th {IEEE} International Conference on Software Quality, Reliability and Security, {QRS} 2019, Sofia, Bulgaria, July 22-26, 2019}, 248--259. {IEEE}.

\bibitem[{Lamkanfi et~al.(2010)Lamkanfi, Demeyer, Giger, and Goethals}]{DBLP:conf/msr/LamkanfiDGG10}
Lamkanfi, A.; Demeyer, S.; Giger, E.; and Goethals, B. 2010.
\newblock Predicting the severity of a reported bug.
\newblock In Whitehead, J.; and Zimmermann, T., eds., \emph{Proceedings of the 7th International Working Conference on Mining Software Repositories, {MSR} 2010 (Co-located with ICSE), Cape Town, South Africa, May 2-3, 2010, Proceedings}, 1--10. {IEEE} Computer Society.

\bibitem[{Le and Babar(2022)}]{DBLP:conf/msr/LeB22}
Le, T. H.~M.; and Babar, M.~A. 2022.
\newblock On the Use of Fine-grained Vulnerable Code Statements for Software Vulnerability Assessment Models.
\newblock In \emph{19th {IEEE/ACM} International Conference on Mining Software Repositories, {MSR} 2022, Pittsburgh, PA, USA, May 23-24, 2022}, 621--633. {ACM}.

\bibitem[{Le, Chen, and Babar(2023)}]{DBLP:journals/csur/LeCB23}
Le, T. H.~M.; Chen, H.; and Babar, M.~A. 2023.
\newblock A Survey on Data-driven Software Vulnerability Assessment and Prioritization.
\newblock \emph{{ACM} Comput. Surv.}, 55(5): 100:1--100:39.

\bibitem[{Le et~al.(2021)Le, Hin, Croft, and Babar}]{DBLP:conf/kbse/LeHCB21/DeepCVA}
Le, T. H.~M.; Hin, D.; Croft, R.; and Babar, M.~A. 2021.
\newblock DeepCVA: Automated Commit-level Vulnerability Assessment with Deep Multi-task Learning.
\newblock In \emph{36th {IEEE/ACM} International Conference on Automated Software Engineering, {ASE} 2021, Melbourne, Australia, November 15-19, 2021}, 717--729. {IEEE}.

\bibitem[{Le, Sabir, and Babar(2019)}]{DBLP:conf/msr/LeSB19}
Le, T. H.~M.; Sabir, B.; and Babar, M.~A. 2019.
\newblock Automated software vulnerability assessment with concept drift.
\newblock In Storey, M.~D.; Adams, B.; and Haiduc, S., eds., \emph{Proceedings of the 16th International Conference on Mining Software Repositories, {MSR} 2019, 26-27 May 2019, Montreal, Canada}, 371--382. {IEEE} / {ACM}.

\bibitem[{Li et~al.(2023{\natexlab{a}})Li, Hammoud, Itani, Khizbullin, and Ghanem}]{DBLP:journals/corr/abs-2303-17760}
Li, G.; Hammoud, H. A. A.~K.; Itani, H.; Khizbullin, D.; and Ghanem, B. 2023{\natexlab{a}}.
\newblock {CAMEL:} Communicative Agents for "Mind" Exploration of Large Scale Language Model Society.
\newblock \emph{CoRR}, abs/2303.17760.

\bibitem[{Li et~al.(2023{\natexlab{b}})Li, Yadavally, Zhang, Wang, and Nguyen}]{DBLP:conf/sigsoft/LiYZWN23}
Li, Y.; Yadavally, A.; Zhang, J.; Wang, S.; and Nguyen, T.~N. 2023{\natexlab{b}}.
\newblock Commit-Level, Neural Vulnerability Detection and Assessment.
\newblock In Chandra, S.; Blincoe, K.; and Tonella, P., eds., \emph{Proceedings of the 31st {ACM} Joint European Software Engineering Conference and Symposium on the Foundations of Software Engineering, {ESEC/FSE} 2023, San Francisco, CA, USA, December 3-9, 2023}, 1024--1036. {ACM}.

\bibitem[{Liang et~al.(2023)Liang, He, Jiao, Wang, Wang, Wang, Yang, Tu, and Shi}]{DBLP:journals/corr/abs-2305-19118}
Liang, T.; He, Z.; Jiao, W.; Wang, X.; Wang, Y.; Wang, R.; Yang, Y.; Tu, Z.; and Shi, S. 2023.
\newblock Encouraging Divergent Thinking in Large Language Models through Multi-Agent Debate.
\newblock \emph{CoRR}, abs/2305.19118.

\bibitem[{NIST(2024)}]{NVD}
NIST. 2024.
\newblock “{National Vulnerability Database (NVD)}”.
\newblock \url{https://nvd.nist.gov/}.

\bibitem[{Ognawala et~al.(2018)Ognawala, Amato, Pretschner, and Kulkarni}]{DBLP:conf/kbse/OgnawalaAPK18}
Ognawala, S.; Amato, R.~N.; Pretschner, A.; and Kulkarni, P. 2018.
\newblock Automatically assessing vulnerabilities discovered by compositional analysis.
\newblock In Perrouin, G.; Acher, M.; Cordy, M.; and Devroey, X., eds., \emph{Proceedings of the 1st International Workshop on Machine Learning and Software Engineering in Symbiosis, MASES@ASE 2018, Montpellier, France, September 3, 2018}, 16--25. {ACM}.

\bibitem[{OpenAI(2023)}]{GPT4}
OpenAI. 2023.
\newblock {GPT-4} Technical Report.
\newblock \emph{CoRR}, abs/2303.08774.

\bibitem[{Peng et~al.(2023)Peng, Wang, Wang, Gao, and Lyu}]{DBLP:conf/kbse/PengWWGL23}
Peng, Y.; Wang, C.; Wang, W.; Gao, C.; and Lyu, M.~R. 2023.
\newblock Generative Type Inference for Python.
\newblock In \emph{38th {IEEE/ACM} International Conference on Automated Software Engineering, {ASE} 2023, Luxembourg, September 11-15, 2023}, 988--999. {IEEE}.

\bibitem[{Sawadogo et~al.(2021)Sawadogo, Guimard, Bissyand{\'{e}}, Kabor{\'{e}}, Klein, and Moha}]{DBLP:journals/corr/abs-2112-10123}
Sawadogo, A.~D.; Guimard, Q.; Bissyand{\'{e}}, T.~F.; Kabor{\'{e}}, A.~K.; Klein, J.; and Moha, N. 2021.
\newblock Early Detection of Security-Relevant Bug Reports using Machine Learning: How Far Are We?
\newblock \emph{CoRR}, abs/2112.10123.

\bibitem[{Smyth(2017)}]{DBLP:journals/ns/Smyth17}
Smyth, V. 2017.
\newblock Software vulnerability management: how intelligence helps reduce the risk.
\newblock \emph{Netw. Secur.}, 2017(3): 10--12.

\bibitem[{Spanos and Angelis(2018)}]{DBLP:journals/jss/SpanosA18}
Spanos, G.; and Angelis, L. 2018.
\newblock A multi-target approach to estimate software vulnerability characteristics and severity scores.
\newblock \emph{J. Syst. Softw.}, 146: 152--166.

\bibitem[{Statista(2024)}]{Statista1}
Statista. 2024.
\newblock Number of common IT security vulnerabilities and exposures (CVEs) worldwide from 2009 to 2024 YTD.
\newblock \url{https://www.statista.com/statistics/500755/worldwide-common-vulnerabilities-and-exposures/}.

\bibitem[{Thongtanunam et~al.(2015)Thongtanunam, McIntosh, Hassan, and Iida}]{DBLP:conf/msr/ThongtanunamMHI15}
Thongtanunam, P.; McIntosh, S.; Hassan, A.~E.; and Iida, H. 2015.
\newblock Investigating Code Review Practices in Defective Files: An Empirical Study of the Qt System.
\newblock In Penta, M.~D.; Pinzger, M.; and Robbes, R., eds., \emph{12th {IEEE/ACM} Working Conference on Mining Software Repositories, {MSR} 2015, Florence, Italy, May 16-17, 2015}, 168--179. {IEEE} Computer Society.

\bibitem[{Thung et~al.(2012)Thung, Lo, Jiang, Lucia, Rahman, and Devanbu}]{DBLP:conf/icsm/ThungLJLRD12}
Thung, F.; Lo, D.; Jiang, L.; Lucia; Rahman, F.; and Devanbu, P.~T. 2012.
\newblock When would this bug get reported?
\newblock In \emph{28th {IEEE} International Conference on Software Maintenance, {ICSM} 2012, Trento, Italy, September 23-28, 2012}, 420--429. {IEEE} Computer Society.

\bibitem[{Wang et~al.(2019)Wang, Zhou, Sun, and Zhang}]{DBLP:conf/icaci/WangZSZ19}
Wang, P.; Zhou, Y.; Sun, B.; and Zhang, W. 2019.
\newblock Intelligent Prediction of Vulnerability Severity Level Based on Text Mining and XGBboost.
\newblock In \emph{Eleventh International Conference on Advanced Computational Intelligence, {ICACI} 2019, Guilin, China, June 7-9, 2019}, 72--77. {IEEE}.

\bibitem[{{WhiteSource}(2023)}]{Mend}
{WhiteSource}. 2023.
\newblock “{M}end bolt”.
\newblock \url{https://www.mend.io/free-developer-tools/}.

\bibitem[{Yamamoto, Miyamoto, and Nakayama(2015)}]{DBLP:conf/badgers/YamamotoMN15}
Yamamoto, Y.; Miyamoto, D.; and Nakayama, M. 2015.
\newblock Text-Mining Approach for Estimating Vulnerability Score.
\newblock In \emph{4th International Workshop on Building Analysis Datasets and Gathering Experience Returns for Security, BADGERS@RAID 2015, Kyoto, Japan, November 5, 2015}, 67--73. {IEEE}.

\bibitem[{Yin, Ni, and Wang(2024{\natexlab{a}})}]{DBLP:journals/corr/abs-2404-02056/llmsva}
Yin, X.; Ni, C.; and Wang, S. 2024{\natexlab{a}}.
\newblock Multitask-based Evaluation of Open-Source {LLM} on Software Vulnerability.
\newblock \emph{CoRR}, abs/2404.02056.

\bibitem[{Yin, Ni, and Wang(2024{\natexlab{b}})}]{DBLP:journals/corr/abs-2404-02056}
Yin, X.; Ni, C.; and Wang, S. 2024{\natexlab{b}}.
\newblock Multitask-based Evaluation of Open-Source {LLM} on Software Vulnerability.
\newblock \emph{CoRR}, abs/2404.02056.

\bibitem[{Zheng et~al.(2023)Zheng, Chiang, Sheng, Zhuang, Wu, Zhuang, Lin, Li, Li, Xing, Zhang, Gonzalez, and Stoica}]{DBLP:conf/nips/ZhengC00WZL0LXZ23}
Zheng, L.; Chiang, W.; Sheng, Y.; Zhuang, S.; Wu, Z.; Zhuang, Y.; Lin, Z.; Li, Z.; Li, D.; Xing, E.~P.; Zhang, H.; Gonzalez, J.~E.; and Stoica, I. 2023.
\newblock Judging LLM-as-a-Judge with MT-Bench and Chatbot Arena.
\newblock In Oh, A.; Naumann, T.; Globerson, A.; Saenko, K.; Hardt, M.; and Levine, S., eds., \emph{Advances in Neural Information Processing Systems 36: Annual Conference on Neural Information Processing Systems 2023, NeurIPS 2023, New Orleans, LA, USA, December 10 - 16, 2023}.

\bibitem[{Zheng et~al.(2021)Zheng, Pujar, Lewis, Buratti, Epstein, Yang, Laredo, Morari, and Su}]{DBLP:conf/icse/ZhengPLBEYLMS21}
Zheng, Y.; Pujar, S.; Lewis, B.~L.; Buratti, L.; Epstein, E.~A.; Yang, B.; Laredo, J.; Morari, A.; and Su, Z. 2021.
\newblock {D2A:} {A} Dataset Built for AI-Based Vulnerability Detection Methods Using Differential Analysis.
\newblock In \emph{43rd {IEEE/ACM} International Conference on Software Engineering: Software Engineering in Practice, {ICSE} {(SEIP)} 2021, Madrid, Spain, May 25-28, 2021}, 111--120. {IEEE}.

\bibitem[{Zhou et~al.(2021)Zhou, Pacheco, Wan, Xia, Lo, Wang, and Hassan}]{DBLP:conf/kbse/ZhouPW00WH21}
Zhou, J.; Pacheco, M.; Wan, Z.; Xia, X.; Lo, D.; Wang, Y.; and Hassan, A.~E. 2021.
\newblock Finding {A} Needle in a Haystack: Automated Mining of Silent Vulnerability Fixes.
\newblock In \emph{36th {IEEE/ACM} International Conference on Automated Software Engineering, {ASE} 2021, Melbourne, Australia, November 15-19, 2021}, 705--716. {IEEE}.

\bibitem[{Zhou et~al.(2022)Zhou, Siow, Wang, Liu, and Liu}]{DBLP:journals/tosem/ZhouSWLL22}
Zhou, Y.; Siow, J.~K.; Wang, C.; Liu, S.; and Liu, Y. 2022.
\newblock {SPI:} Automated Identification of Security Patches via Commits.
\newblock \emph{{ACM} Trans. Softw. Eng. Methodol.}, 31(1): 13:1--13:27.

\end{thebibliography}

\end{document}